\documentclass[twocolumn,showpacs,preprintnumbers,amsmath,prd,amssymb]{revtex4-2}
\usepackage{natbib}
\usepackage{multirow}
\usepackage{graphicx,subfigure}
\usepackage{bm}
\usepackage{amssymb,amsmath,amsfonts,latexsym}
\usepackage[colorlinks=true,linktocpage=true,linkcolor=blue,citecolor=blue]{hyperref}
\usepackage[utf8]{inputenc}
\usepackage{lipsum}
\usepackage{slashed}
\usepackage{mathtools}
\usepackage{caption}
\usepackage{lineno}
\usepackage{orcidlink}
\usepackage{subfiles} 
\usepackage{xcolor}
\def\bs{\boldsymbol} 

\def\bdel{\bs\partial}

\long\def\comment#1{ }

\def\and{\qquad\text{and}\qquad}

\def\0{{\boldsymbol 0}}
\def\q{{\boldsymbol q}}

\def\x{{\boldsymbol x}}
\def\y{{\boldsymbol y}}
\def\p{{\boldsymbol p}}

\def\z{{\boldsymbol z}}

\def\Tan{\text{Tan}}
\def\tm{t_{m}}
\def\t0{t_{0}}
\def\O0{\Omega_{0}}
\def\HO{\text{\tiny HO}}

\def\pT{p_{_T}}

\def\Kc{{\cal K}}
\def\cK{{\cal K}}
\newcommand{\beq}{\begin{eqnarray}}
\newcommand{\eeq}{\end{eqnarray}}
\newcommand{\be}{\begin{eqnarray*}}
\newcommand{\ee}{\end{eqnarray*}}
\newcommand{\bal}{\begin{align}}
\newcommand{\eal}{\end{align}}
\newcommand{\rmd}{{\rm d}}

\newcommand{\rme}{{\rm e}}

\def\bdel{\bs \partial}
\def\bs{\boldsymbol} 

\def\bdel{\bs\partial}
\long\def\comment#1{ }

\def\and{\qquad\text{and}\qquad}

\def\0{{\boldsymbol 0}}
\def\q{{\boldsymbol q}}

\def\x{{\boldsymbol x}}
\def\y{{\boldsymbol y}}
\def\p{{\boldsymbol p}}

\def\z{{\boldsymbol z}}

\def\Tan{\text{Tan}}
\def\tm{t_{m}}
\def\t0{t_{0}}
\def\O0{\Omega_{0}}
\def\HO{\text{\tiny HO}}

\def\pT{p_{_T}}

\def\Kc{{\cal K}}
\def\cK{{\cal K}}

\def\bdel{\bs \partial}
\newcommand{\nn}{\nonumber\\ }

\begin{document}
\title{Sensitivity of jet quenching to the initial state in heavy-ion collisions}
\author{Souvik Priyam Adhya\orcidlink{https://orcid.org/0000-0002-4825-2827}} 
\email{souvikadhya2007@gmail.com (corresponding author)}
\affiliation{Institute of Physics of the Czech Academy of Sciences, Na Slovance 1999/2, 18221 Prague 8, Czech Republic}
\author{Konrad Tywoniuk\orcidlink{https://orcid.org/0000-0001-5677-0010}}
\email{konrad.tywoniuk@uib.no (corresponding author)}
\affiliation{Department of Physics and Technology, University of Bergen, 5007 Bergen, Norway}%
\begin{abstract}
In heavy-ion collisions, nuclear matter is subjected to extreme conditions in a highly dynamical, rapidly evolving environment. This poses a tremendous challenge for calculating jet quenching observables. Current approaches rely on analytical results for static cases, introducing theoretical uncertainties and biases in our understanding of the pre-equilibrated medium. To address this issue, we employ resummation schemes to derive analytical rates for radiative energy loss in generic, evolving backgrounds. We investigate regimes where rare scattering and multiple scattering with the dynamical medium occurs, and extract relevant scales governing the in-medium emission rate of soft gluons. Our analysis indicates that strong jet quenching is only possible when the equilibration time of the medium is longer than its mean free path, highlighting the importance of medium modifications of jets in the earliest stages of heavy-ion collisions. We also demonstrate analytically that a medium evolution, which initially has a small coupling to jets, typically leads to a stronger jet azimuthal asymmetry at the same jet suppression factor.   
\end{abstract}
\date{\today}
\maketitle
\section{Introduction} 
The quark-gluon plasma (QGP), created in relativistic heavy-ion collisions, is short-lived but leaves a strong imprint on particle production. The focus of this work is on the high-energy particles that pass through the QGP and can be used to study its microscopic properties. These particles are detected as collimated sprays of particles and energy, called jets \cite{Ellis:1996mzs, Dokshitzer:1991wu}. Jet modifications \cite{Gyulassy:1990ye,Gyulassy:2003mc,Peigne:2008wu,Mehtar-Tani:2013pia,Ghiglieri:2015zma,Andrews:2018jcm}, compared to a baseline obtained in proton-proton collisions, have been studied extensively for over a decade at RHIC and LHC \cite{STAR:2002svs,STAR:2003pjh,Aad:2010bu,Chatrchyan:2011sx,Aad:2012vca,Abelev:2013kqa,Aad:2014bxa,Chatrchyan:2014ava,Aad:2014wha,Aad:2015bsa,Khachatryan:2016tfj,Khachatryan:2016jfl,Aaboud:2017bzv,Aaboud:2017eww,ATLAS:2018gwx,ALICE:2019qyj,ATLAS:2021ktw,STAR:2021kjt,ATLAS:2022vii,ATLAS:2022zbu,ALICE:2022vsz,ALICE:2023jye,ATLAS:2023hso}.
The magnitude of these modifications is quantified by the suppression factor, $R_{AA}$, which is the ratio of the jet yield in heavy ion collisions to that in proton-proton collisions scaled by the number of binary nucleon-nucleon collisions, and the dependence on medium geometry is probed through the azimuthal flow harmonic coefficient $v_2$ of jets. In addition, other jet observables, such as jet shapes, fragmentation patterns, and substructure measurements, have also been studied 
\cite{Armesto:2015ioy,Connors:2017ptx,Kogler:2018hem,Cunqueiro:2021wls,Apolinario:2022vzg}. These observables provide complementary information on the phenomenon of jet quenching to investigate the properties of the QGP.

\begin{figure}
    \centering
    \includegraphics[width=\columnwidth]{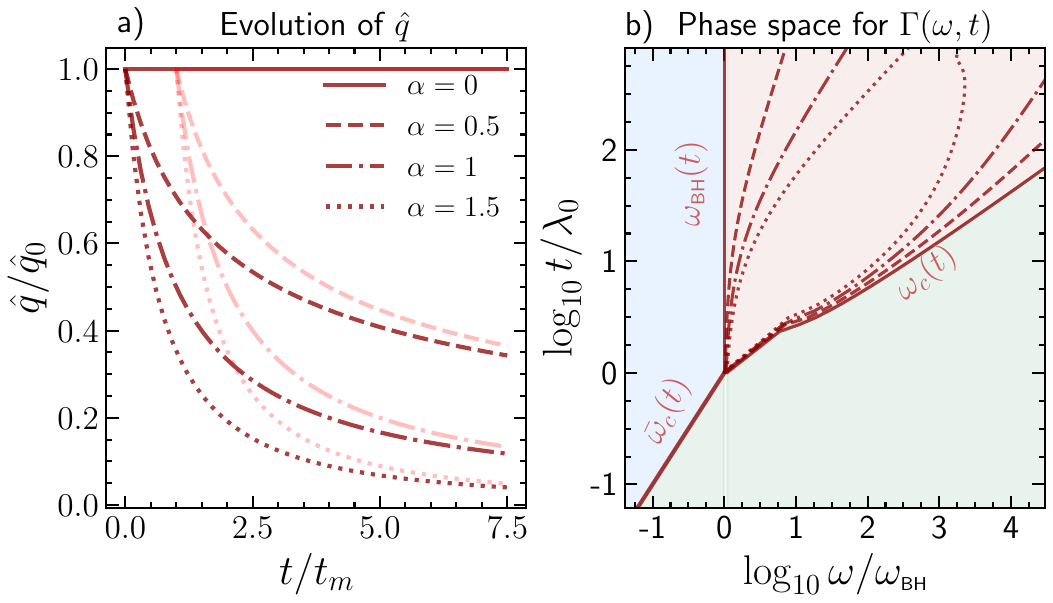}
    \caption{a) Two models for the evolution of $\hat q$, model \texttt{(i)} represents the equilibriation of an initially over-occupied system, while \texttt{(ii)} corresponds to an initially under-occupied system. b) Evolution of the relevant energy scales governing the in-medium emission rate $\Gamma(\omega,t)$.}
    \label{fig:fig1}
\end{figure}

The produced jets are moving through a constantly evolving and diluting background. Jet quenching observables therefore capture both the integrated and differential properties of medium evolution, making them sensitive to various stages of heavy-ion collisions. At late times, the bulk matter achieves a state of ``hydrodynamic'' equilibrium, meaning that its subsequent evolution can be captured by nearly-perfect hydrodynamics \cite{Song:2010mg}. In the Bjorken model \cite{Bjorken:1982qr}, representing boost-invariant relativistic hydrodynamics, the QGP energy density $\varepsilon$ evolves according to
\begin{equation}
    \varepsilon(t) = \varepsilon_0\left(\frac{t_0}{t} \right)^{1+c_s^2} \,,
\end{equation}
where $\varepsilon_0$ is the initial energy density, $t_0$ is the initial time of the hydrodynamical evolution and $c_s^2 = \partial P/\partial \varepsilon$ is the speed of sound. In an ideal, relativistic plasma, $c_s^2 = 1/3$. Since the jet transport parameter behaves like $\hat q \propto g^4 \varepsilon^{3/4} \sim t^{-1}$, describing both the momentum broadening of propagating particles and governs energy loss in the medium, we should expect that the jet quenching effects gradually weaken during the hydrodynamic expansion of the QGP \cite{Adhya:2019qse,Caucal:2020xad,Adhya:2021kws,Adhya:2022tcn}. In this work, we consider a class of generalized models with $\hat q \sim t^{-\alpha}$, see Fig.~\ref{fig:fig1}a).

How the hot and dense medium evolves prior to the equilibration time is largely an open question. For an initially over-occupied system of gluons one expects the so-called ``bottom-up'' equilibration scenario \cite{Mueller:1999pi,Baier:2000sb,Berges:2013eia,Berges:2013fga,Kurkela:2015qoa} that indicates a rapid growth of the momentum broadening of hard partons followed by a gradual decay \cite{Ipp:2020mjc,Ipp:2020nfu,Carrington:2021dvw,Carrington:2022bnv,Avramescu:2023qvv,Barata:2024xwy}. In kinetic theory, $\hat q$ has been tracked from the early ``bottom-up'' stage to late-time hydro evolution \cite{Boguslavski:2023alu,Boguslavski:2023waw}, finding evidence of a limiting attractor \cite{Boguslavski:2023alu}, see also \cite{Boguslavski:2024kbd}. As an illustrative example \cite{Carrington:2021dvw}, $\hat q \approx 6$ GeV$^2$/fm at the very early stages of the collision $\tau \sim 0.06$ fm \footnote{In most works on hard probes in glasma, one actually studies the momentum broadening of particles. For a discussion of the modified radiative spectrum, see e.g. \cite{Barata:2024xwy}.}. 
In contrast, when starting in an under-occupied partonic system there is a certain delay before interactions overcome the rapid longitudinal expansion of the system, causing $\hat q$ initially to grow before reaching a maximum value, subsequently followed by a hydrodynamic decay. The latter scenario is appealing from a phenomenological point of since it gives rise to an enhancement of the azimuthal asymmetry of the jet suppression \cite{Andres:2020vxs,Andres:2022bql}, encoded in the $v_2$ parameter to be discussed below. 

In this work, we assess the potential of jet quenching observables to capture the imprint of early dynamics in heavy-ion collisions. To accomplish this, we propose two distinct scenarios:
\begin{itemize}
    \item Model \texttt{(i)} aims at capturing the characteristics of an initially over-occupied system, featuring an initial large value of $\hat q$ with a subsequent decay. 
    \item In contrast, model \texttt{(ii)} allows for a delayed onset of quenching effects up to a thermalization time $\tm$. 
\end{itemize}
The two models are more rigorously defined in Eq.~\eqref{eq:qhat-models} below. To allow for flexibility, $\tm$ is independent of the mean free path of elastic scattering $\lambda$. In both cases we consider a flexible parameterization of the hydro-like regime, $\hat q \sim t^{-\alpha}$, see Fig.~\ref{fig:fig1}a) for details.

In order to tackle these questions, we first derive analytical in-medium emission rates to calculate energy loss that are valid for any medium expansion scenario, see \cite{Huss:2020whe} for  discussion on small systems. This can be achieved by using established schemes to deal with scattering on a background medium \cite{Isaksen:2022pkj}, including the improved opacity expansion (IOE) approach \cite{Mehtar-Tani:2019tvy,Mehtar-Tani:2019ygg,Barata:2020sav,Barata:2020rdn,Barata:2021wuf}. While earlier studies assumed a static medium, this letter extends the analysis to include the relevant energy scales that govern radiative energy loss in an expanding medium. This allows us to determine the validity of the multiple, soft and rare, hard scatterings not only as a function of energy but also as a function of the initial quenching time of the medium, see Fig.~\ref{fig:fig1}b). 

This novel approach allows us to determine whether multiple scatterings play a significant role in dynamical systems, which in turn influences the design of radiative in-medium parton showers.

\section{Formalism} 
\label{sec:formalism}

\subsection{Generic expressions of medium induced rate}
\label{sec:rates}

We study the medium-induced spectrum for emitting soft gluons, $\Gamma(\omega,t) = \rmd I^\text{\tiny med}/(\rmd \omega \rmd t)$ \cite{Baier:1996kr,Zakharov:1997uu,Arnold:2002ja}, which reads \cite{Dominguez:2012ad,Isaksen:2022pkj}
\begin{align}
	\label{eq:med-rate}
	\Gamma = \frac{4 \alpha_s C_{\scriptscriptstyle R}}{\omega^2} \int_0^{t} \rmd t_0 \int_{\p,\p_0} \!\!\Sigma(\p^2,t) \frac{\p\cdot \p_0}{\p^2}  \tilde\Kc(\p,t; \p_0,t_0) ,
\end{align}
where $C_R$ is the color charge of the emitting particle and $\Sigma(\p^2,t)  =  \int_\q \, \Theta(\q^2-\p^2) \sigma(\q,t)$, with the scattering potential off gluons in the medium being encoded in
\begin{equation}
    \sigma(\q,t) = N_c n(t) \frac{\rmd\sigma_\text{\tiny el}}{\rmd^2\q} \,,
\end{equation}
where $n(t)$ denotes the density of scattering centers \footnote{We have also introduced the notation for transverse integrals, where $\int_\q = \int \rmd^2q/(2\pi)^2$ and $\int_\x = \int \rmd^2 x$.}. 
We will also define the (time-dependent) mean free path as $\lambda(t) \equiv \Sigma^{-1}(0,t)$. Finally, with $\tilde \Kc = (-{\rm Im }) \Kc$, the three-point correlator $\Kc(\p,t;\p_0,t_0)$ describes the transverse momentum broadening experienced by the gluon during its formation time, and is the Green's function of a Schr\"odinger equation in transverse-coordinate space, i.e. 
\begin{equation}
    \left[i\partial_t + \frac{{\bs \partial}_\perp^2}{2\omega} + i v(\x,t) \right] \Kc(\x,t;\y,t_0) = i \delta(\x - \y)\delta(t-t_0) \,,
\end{equation}
where the time-dependent medium potential $v(\x,t)$ is related to the elastic scattering cross section through $v(\x,t) = \int_\q \sigma(\q,t) \big(1-\rme^{i \q \cdot \x} \big)$.

The key quantity of the problem is the so-called jet quenching parameter 
\begin{equation}
    \hat q_0 = 4\pi \alpha_s^2 N_c n(t) \,,
\end{equation}
which is a direct measure of the scattering density. In the widely used Gyulassy-Wang (GW) model for medium interactions, 
\begin{equation}
    \sigma(\q,t) = \frac{4\pi \hat q_0(t)}{(\q^2+\mu^2)^2} \,,
\end{equation}
where $\mu$ is a medium screening mass \cite{Wang:1992qdg}. This yields $\Sigma(\p^2,t) = \hat q_0(t)/(\p^2+\mu^2)$. The mean-free-path of this medium is then given by  $\lambda(t) = 1/\Sigma(0,t) = \mu^2/\hat q_0(t)$. The temperature dependence mainly affects the density of scattering centers, $n \propto T^3$ \footnote{In a thermal medium, the in-medium screening mass is temperature dependent, $\mu^2 \propto T^2$, but we will neglect this effect since it only gives rise to logarithmic corrections for emissions at the thermal scale, $\omega \sim T$, and comment on the expected effects below.}. In this section we derive completely generic expressions for the rate of medium-induced emissions, valid for \emph{any} medium evolution model. The evolution of the relevant scales of the radiative process is summarized in Fig.~\ref{fig:fig1}b).

Performing a full resummation of multiple interactions can be done employing numerical \cite{Zakharov:2004vm,CaronHuot:2010bp,Feal:2018sml,Ke:2018jem,Andres:2020vxs,Schlichting:2020lef} or analytical techniques. For sufficiently dilute or small media, the probability of one scattering dominates \cite{Gyulassy:1999zd,Wiedemann:2000za,Sievert:2018imd}. On the contrary, for large or dense media, multiple scattering on the background medium start to play a role \cite{Baier:1996kr,Zakharov:1997uu,Salgado:2003gb,Armesto:2003jh}. In this case, there will be an interplay between copious, soft interactions and rare, hard momentum exchanges. The soft regime is governed by LPM interference effects \cite{LPM1, LPM2, Migdal:1956tc}, allowing for the possibility of rapid energy-degradation of the emitting particle. It is therefore of utmost importance to establish whether this regime is relevant for expanding media, where the density rapidly decays with proper time. In order to address this question, we will employ three resummation schemes that, in total, cover the whole emission phase space in gluon energy and time $(\omega, t)$ \cite{Isaksen:2022pkj}.

\subsubsection{Opacity expansion}
Early in the jet propagation, when the time is shorter than the local mean-free-path $t \ll \lambda(t)$, the probability of interacting with the medium is still small and we can truncate the conventional \textit{opacity expansion} (OE) at first order. 

In the opacity expansion (OE), the 3-point function with support in the time domain $(t_0,t)$ is expanded around the vacuum solution, given as Dyson-type implicit equation,
\begin{equation}
    \begin{split}
    \label{eq:OE-master}
        \Kc(\p;\p_0) &= (2\pi)^2 \delta(\p-\p_0)\Kc_0(\p) \\ 
        &-\int \rmd s \int_\q \, \Kc_0(\p) v(\q,s) \Kc(\p-\q;\p_0) \,,
    \end{split}
\end{equation}
where the interaction $v(\q) = (2\pi)^2 \delta(\q)\Sigma(0,s) - \sigma(\q,s)$ occurs at some intermediate time $t_0 < s < t$ and we have suppressed the time-variables for simplicity. The vacuum propagator is given by the plane-wave solution $\Kc_0(\p) = \exp[-i\frac{\p^2}{2\omega}(t-t_0)]$.

Since Eq.~\eqref{eq:med-rate} is a \textit{medium-induced} rate, i.e. has no contributions from vacuum, it suffices to evaluate the three-point function $\Kc(\p;\p_0)$ at zeroth order in the expansion \eqref{eq:OE-master}. The rate at first order in opacity can then be written as
\begin{equation}
\label{eq:rate-oe}
\Gamma_\text{\tiny OE}(\omega,t) = \frac{2\bar \alpha}{\lambda(t)\, \omega} \mathcal{F}\left(\frac{\bar \omega_c(t)}{\omega}\right)\,,
\end{equation}
where $\mathcal{F}(x) = \ln(x) + \gamma_E - \cos(x){\rm Ci}(x) + \sin(x) \big(\frac{\pi}{2} - {\rm Si}(x) \big)$, with ${\rm Ci}(x)$ and ${\rm Si}(x)$ the cosine and sine integral functions. The energy scale is found to be $\bar \omega_c(t) = \mu^2 t/2$.
This is immediately implies the limiting behavior for the rate, as
\begin{equation}
\label{eq:rate-oe-limits}
\Gamma_\text{\tiny OE}(\omega,t) = \frac{2 \bar \alpha}{\lambda(t)\,\omega} \begin{cases} \ln\frac{\bar \omega_c(t)}{\omega} + \gamma_E&  \text{for } \omega \ll \bar \omega_c(t) \,, \\ \frac{\pi}2 \frac{\bar \omega_c(t)}{\omega} & \text{for } \omega \gg \bar \omega_c(t) \,. \end{cases}
\end{equation}

\subsubsection{Resummed opacity expansion}

At later times, $t \gg \lambda(t)$, multiple interactions have to be taken into account. Formally, this is manifested by divergences for soft gluon emissions at arbitrary orders in the OE \cite{Isaksen:2022pkj}. 

Since other resummation schemes are called for, we first derive the rate at first order in the \textit{resummed} opacity expansion (ROE). In this scheme, one expands the 3-point function as \begin{equation}
    \begin{split}
    \label{eq:ROE-master}
        & \Kc(\p;\p_0) = (2\pi)^2 \delta(\p-\p_0)\Delta(t,t_0)\Kc_0(\p) \\
        & - \int \rmd s \int_\q\, \Delta(t,s)\Kc_0(\p) \sigma(\q,s) \Kc(\p-\q;\p_0) \,,
    \end{split}
\end{equation}
see \cite{Isaksen:2022pkj}. In contrast to \eqref{eq:OE-master} this expansion involves only the scattering potential $\sigma(\q,s)$. In addition, a novel element is the elastic Sudakov form factor $\Delta(t,t_0)$, that for a generic medium profile can be written as,
\begin{equation}
    \label{eq:ROE-sudakov}
     \Delta(t,t_0) \equiv \exp\left[-\int_{t_0}^{t} \frac{\rmd s}{\lambda(s)} \right] \,.
\end{equation}
At first order in this expansion the ROE spectrum only the zeroth order of Eq.~\eqref{eq:ROE-master} is needed, giving
\begin{equation}
\label{eq:ROE-rate}
    \Gamma_\text{\tiny ROE}(\omega,t) = \bar \alpha \frac{\hat q_0(t) t}{\omega^2} \int_0^1 \rmd s \, \Delta(t,st)f\big(u(1-s) \big) \,,
\end{equation}
where $u = \bar \omega_c(t)/\omega$ and $f(x) = \sin(x){\rm Ci}(x) + \cos(x)\big(\frac{\pi}2 - {\rm Si}(x) \big)$. Given that the Sudakov factor is exponentially decaying, we can approximate $\Delta(t,t_0) \simeq \rme^{-(t-t_0)/\lambda(t)}$. 
The asymptotic limits can easily be obtained as,
\begin{equation}
\label{eq:rate-roe-limits}
\Gamma_\text{\tiny ROE} \simeq \frac{2\bar \alpha}{\lambda(t)\omega} \begin{cases} \ln \frac{\omega_\text{\tiny BH}(t)}{\omega} +\text{Ei}\left(-\frac{t}{\lambda(t)}\right) & \text{for } \omega \to 0 \,,\\ \frac{\pi}2 \frac{\omega_\text{\tiny BH}(t)}{\omega} \left(1-\rme^{-t/\lambda(t)} \right) & \text{for } \omega \to \infty \,,  \end{cases}
\end{equation}
where $\omega_\text{\tiny BH}(t) = \mu^2 \lambda(t)/2$.
In the dilute case, $t \ll \lambda(t)$, we directly recover the OE results in Eq.~\eqref{eq:rate-oe-limits} where the scale becomes $\bar\omega_c(t)$.

In the regime of multiple scattering, $t \gg \lambda(t)$, due to the transmutation of scales, we only relevant energy becomes $\omega_\text{\tiny BH}(t)$. We get
\begin{equation}
\label{eq:rate-roe-limits-dense}
\Gamma_\text{\tiny ROE} (\omega,t) \simeq \frac{2\bar \alpha}{\lambda(t)\,\omega} \begin{cases} \ln \frac{\omega_\text{\tiny BH}(t)}{\omega} & \text{for } \omega \ll \omega_\text{\tiny BH}(t) \,,\\ \frac{\pi}2 \frac{\omega_\text{\tiny BH}(t)}{\omega}  & \text{for } \omega \gg \omega_\text{\tiny BH}(t) \,.  \end{cases}
\end{equation}

\subsubsection{Improved opacity expansion}
For emissions with $\omega > \omega_\text{\tiny BH}(t)$ in the multiple-scattering regime, the so-called \textit{improved opacity expansion} (IOE) is the most appropriate \cite{Mehtar-Tani:2019tvy,Mehtar-Tani:2019ygg,Barata:2020sav,Barata:2020rdn,Barata:2021wuf,Isaksen:2022pkj}. In this framework, we write the potential as 
\begin{equation}
    \label{eq:IOE-potential}
    v(\x,t) \approx v_\HO(\x,t) + \delta v (\x,t) \,,
\end{equation}
where $v_\HO(\x,t)= \hat q(t) \x^2/4$ and $\delta v(\x,t) = \hat q_0 (t) \ln(1/(\x^2 Q^2)) \x^2/4$, where $Q^2$ is a separation scale and $\mu_\ast^2 = \mu^2 \rme^{-1+2 \gamma_E}/4$ for the GW model. Here, we also introduced a ``dressed'' jet parameter $\hat q = \hat q_0 \ln Q^2/\mu^2_\ast$ and it's time dependent analogue $\hat q(t) = \hat q_0(t) \ln Q^2/\mu^2_\ast$. The IOE formally is well-posed as long as $Q^2 \gg \mu_\ast^2$. 

It is most convenient to work with the expression for the rate \eqref{eq:med-rate} Fourier transformed to transverse coordinate space, see e.g. \cite[App.~B]{Isaksen:2022pkj}, namely
\begin{equation}
    \begin{split}
        \Gamma(\omega,t) &= \frac{2\bar\alpha}{\omega^2} \int_0^t \rmd t_0 \int_\z \, v(\z,t) \\ 
        & \times \frac{\z}{\z^2}\cdot \bdel_\y (-{\rm Im})\Kc(\z,t;\y,t_0)\big|_{\y=0} \,.
    \end{split}
\end{equation}
Given the separation \eqref{eq:IOE-potential}, the expansion of the 3-point function $\Kc(\x,t;\y,t_0)$ can be cast as an implicit equation,
\begin{equation}
    \begin{split}
    \label{eq:cK_ful_IOE_app}
        \cK(\x;\y) &= \cK_\HO(\x;\y) \\
        &-\int  \rmd s \int_\z \, \cK_\HO(\x;\z) \delta v(\z,s) \cK(\z;\y) \,,  
    \end{split}
\end{equation}
where $\Kc_\HO(\x,\y)$ is the full correlator with the harmonic potential $v_\HO(\x,t)$ and is given in Eq.~\eqref{eq:cK_BDMPS_app}. For an in-depth discussion of the explicit solutions for different medium expansion scenarios we refer to App.~\ref{sec:IOE-solutions}.

Up to the second order in the IOE expansion, we find $\Gamma_\text{\tiny IOE}(\omega,t) =\Gamma_\text{\tiny IOE}^{(0)}(\omega,t) +\Gamma_\text{\tiny IOE}^{(1)}(\omega,t) $, with
\begin{align}
\label{eq:gamma_IOE_0}
\Gamma_\text{\tiny IOE}^{(0)}(\omega,t) &= \frac{\bar \alpha \hat q (t)}{\omega^2} (- {\rm Im}) {\rm Tan}(t)\,,\\
\label{eq:gamma_IOE_1}
\Gamma_\text{\tiny IOE}^{(1)}(\omega,t) &= \frac{\bar \alpha \hat q_0 (t)}{\omega^2} (- {\rm Im}) {\rm Tan}(t)
\,\ln \frac{\omega \, {\rm Cot}(t)}{2i \rme^{-\gamma_E}\,Q^2}  \,,
\end{align}
where the characteristic scale is $\omega_c(t) = \hat q(t) t^2/2$. 
In the next-to-harmonic term $\Gamma^{(1)}_\text{\tiny IOE}(\omega,t)$ we have dropped a non-local term which is of higher order in $\hat q_0$, i.e. $\mathcal{O}(\hat q_0 \sqrt{\hat q_0})$. We subleading contribution is partially compensated by a proper choice of the matching scale, see below.

Here, $\Tan(t) = S(t,0)/C(0,t)$ where $S(t,t_0)$ and $C(t,t_0)$ are the two characteristic solutions of the harmonic equation $[\partial_t^2 +\Omega^2(t)]f(t,t_0) = 0$, where $\Omega(t) = \sqrt{-i \hat q(t)/(2\omega)}$, with appropriate boundary conditions \cite{Arnold:2008iy}, see App.~\ref{sec:IOE-solutions} for further details.

 The soft limit of these rates follows straightforwardly from $\lim_{\omega \to 0} {\rm Tan}(t) = (1-i) \sqrt{\omega/\hat q(t)}$, as
\begin{align}
\lim_{\omega \to 0} \Gamma_\text{\tiny IOE}^{(0)}(\omega,t) &= \bar \alpha \sqrt{\frac{ \hat q (t)}{\omega^3}} \,,\\
\lim_{\omega \to 0} \Gamma_\text{\tiny IOE}^{(1)}(\omega,t) &= \bar \alpha \frac{ \hat q_0 (t)}{\sqrt{\hat q(t) \omega^3}} \nonumber \\ &\times \left[\ln \frac{\sqrt{\hat q(t) \omega}}{Q^2} + \gamma_E -\frac32 \ln 2 + \frac{\pi}4 \right] \,.
\end{align}
In the hard limit, we obtain $\lim_{\omega \to \infty}{\rm Tan}(t) = t + \mathcal{O}(1/\omega)$. The harmonic oscillator term therefore is strongly suppressed, $\Gamma_\text{\tiny IOE}^{(0)} \sim \mathcal{O}(1/\omega^3)$, and the leading behavior is contained in the second term which, using $\ln i = i \pi/2$, simply reads
\begin{equation}
\label{eq:IOE-rate-hard}
    \lim_{\omega \to \infty} \Gamma_\text{\tiny IOE}^{(1)}(\omega,t) = \bar \alpha \frac{\pi}2 \frac{\hat q_0(t) t}{\omega^2} \,,
\end{equation}
and is identical to the hard limit of the opacity expansion. The characteristic scale related to this transition reads $\omega_c = \hat q(t) t^2/2$.

In order to proceed, we need to find a suitable matching scale $Q^2$. For expanding media, the soft rate becomes explicitly time-dependent, and we have to fix the scale at the level of the rate rather on the level of the spectrum, as done before \cite{Barata:2020sav}. In the limit of soft gluon emissions, the IOE rate converges to a series in $\hat q_0/\hat q$ where the leading order term is the well-known BDMPS-Z result. In this regime, the rate is quasi-instantaneous and we find that the higher-order (next-to-harmonic) corrections scale as 
\begin{equation}
\label{eq:ioe-rate-ratio-soft-approx}
   \frac{\Gamma_\text{\tiny IOE}^{(1)}}{\Gamma_\text{\tiny IOE}^{(0)}}\Bigg|_{\omega \ll \omega_c(t)} \!\!= \frac{\hat q_0}{\hat q}\left[\gamma_E + \frac\pi4 -\frac32\ln 2 + \ln\frac{\sqrt{\hat q(t) \omega}}{ Q^2} \right] .
\end{equation}
We obtain a time-dependent matching scale, that reads
\begin{equation}
    Q^2(t) = \sqrt{\hat q(t) \omega} \,.
\end{equation}
This is one of the main findings of our work. The rate at next-to-harmonic order therefore becomes in the soft limit
\begin{equation}
    \label{eq:rate-ioe-soft}
    \left.\Gamma_\text{\tiny IOE}(\omega,t)\right|_{\omega\ll \omega_c(t)} = \bar \alpha \sqrt{\frac{\hat q_{\rm eff}(t)}{\omega^3}} \,,
\end{equation}
where the effective jet transport parameter becomes $\hat q_{\rm eff}(t) = \hat q(t) (1 + 0.646/\ln \frac{Q^2(t)}{\mu_\ast^2} + \ldots) $ \footnote{It is worth pointing out that our results are slightly different from previous calculations of the IOE rate and spectrum, in particular \cite{Barata:2020sav}. This is due to the missing subleading contribution, see comment below Eq.~\eqref{eq:gamma_IOE_1}.
}. The implicit equation for the matching scale can only be satisfied above some minimal energy which is compatible with the Bethe-Heitler scale, i.e. $\omega > 2\mu_\ast^4\rme/\hat q_0(t)$. Note that the lower boundary can indeed be identified with the Bethe-Heitler energy $\omega_\text{\tiny BH}(t) \sim \mu^4/\hat q_0(t)$. This implies that the IOE should be matched onto the soft limit of the ROE at appropriately soft energies. 

\subsection{Medium expansion models}

\begin{figure}
    \centering
    \includegraphics[width=\columnwidth]{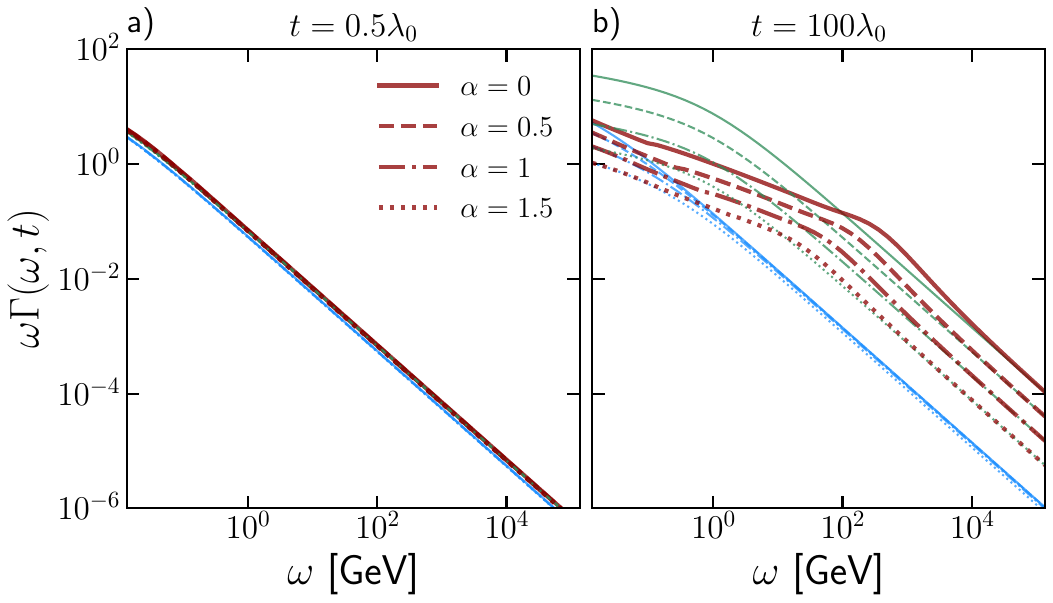}
    \includegraphics[width=\columnwidth]{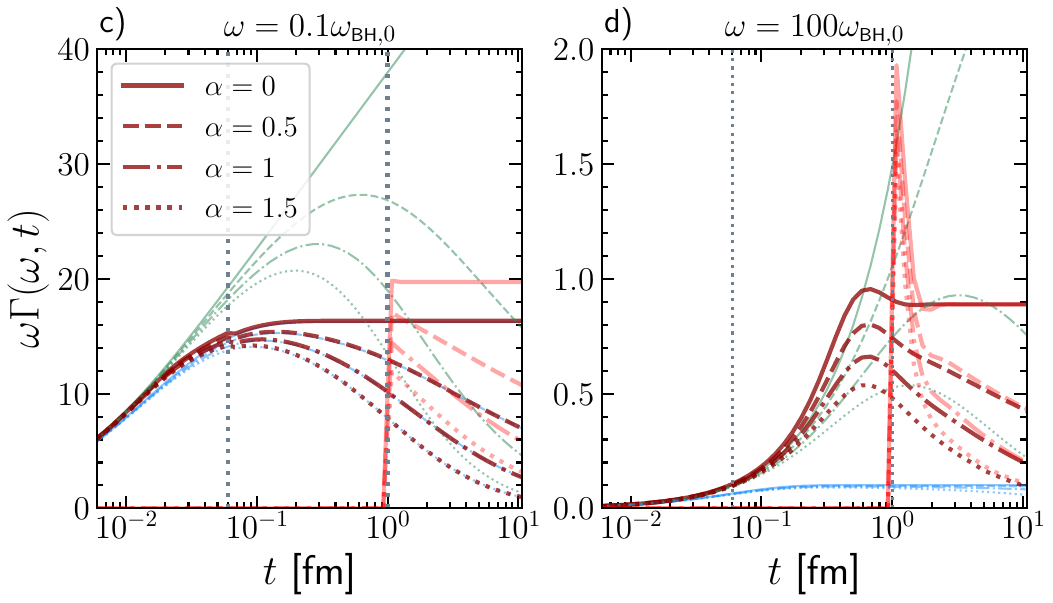}
    \caption{The energy, in a) and b), and time, in c) and d), dependence of $\Gamma(\omega,t)$. Red curves correspond to the full result for model \texttt{(i)}, while green (blue) curves represent the first-order OE and ROE results for reference, see also Fig.~\ref{fig:fig1}. Light-red curves are obtained for model \texttt{(ii)}. Dotted black lines correspond to $\lambda_0$ and $\tm$ values.}
    \label{fig:fig2}
\end{figure}

The rates derived above hold for any medium expansion model. However, in order to interpolate between both of the two cases of medium thermalization described in the introduction, we assume that the temperature behaves parametrically as $T(t) = T_0 [\tm/(t+ \tm)]^{\alpha/3} $. The concrete models under consideration is therefore defined as
\begin{equation}
\label{eq:qhat-models}
\hat q_0(t) = \begin{cases}
    \hat q_0 \left(\frac{\tm}{t + \tm}\right)^\alpha & \text{for model \texttt{(i)}} \,,\\ 
    \hat q_0 \Theta(t-\tm) \left(\frac{\tm}{t} \right)^\alpha & \text{for model \texttt{(ii)}} \,.
\end{cases}
\end{equation}
where $\hat q_0$ defines a constant reference density in the two models, at $t=0$ and $t=t_m$ respectively. A medium diluting according to a 1D Bjorken expansion corresponds to $\alpha=1$, see above, but other cases can also be implemented in realistic phenomenological situations \cite{Andres:2020vxs,Andres:2022bql}.
In this case, in order for $\omega_c(t) \gg \omega_\text{\tiny BH}(t)$, we then have to demand that
 \begin{equation}
    1 \ll \frac{t}{\lambda_0} \left(\frac{\tm} {t}\right)^{\alpha} \,,
 \end{equation}
 where $\lambda_0 = \mu^2/\hat q_0$ is the mean-free-path at $t=0$.
For instance, in the Bjorken scenario, where $\alpha =1$, the medium ``hydrodynamization'' time should be much bigger than the mean-free-path $\tm \gg \lambda$ in order to get contributions from this regime \footnote{The limit $\tm < \lambda_0$ is not of phenomenological relevance as for the hydrodynamic modes to set in the regime, one needs a reasonable mean-free-path to avoid instantaneous interactions. For our parameters $\tm \gg \lambda_0$. In any case, OE captures the dynamics in this regime.}. For our numerical studies in Fig.~\ref{fig:fig2}, we have chosen $\hat q_0=0.3$ GeV$^3$, $\mu^2=0.09$ GeV$^2$, similar to previous studies \cite{Mehtar-Tani:2024jtd}, and $\tm = 1$ fm. We note that for $t<\lambda_0$, corresponding to early emissions in a dilute medium, the medium expansion has no significant effect and the single scattering (OE) rates are valid in that regime. For $t>\lambda_0$, the impact of expansion on both quenching models is substantial across all resummation schemes, which are matched through time-dependent kinematic scales.

The results for model \texttt{(ii)} are easily derived from the explicit results derived for model \texttt{(i)}, merely by taking care of the matching condition to the vacuum at $t<\tm$, see also \cite{Andres:2022bql}. For example, at leading HO order in the IOE, $C(0,L) \mapsto \tilde C(\tm,L) + \tm \partial_L \tilde C(L,\tm)$.
Crucially, Eq.~\eqref{eq:rate-ioe-soft} also holds for model \texttt{(ii)}.

\subsection{Comment on a time-dependent screening mass}

So far we have kept the screening mass constant, but here we briefly comment on the case of a time-dependent screening mass, see e.g. \cite{Andres:2022bql}. Assuming that $\mu $ is related to the Debye screening mass, from LO HTL theory we expect
\begin{equation}
    \mu(t) = \mu_0 \left(\frac{\tm}{t+\tm} \right)^{\alpha/3} \,.
\end{equation}
This effect is straightforwardly implemented in OE and ROE. We can introduce $\lambda'(t)$ as the new mean free path where the time dependent screening mass is taken into account. For the ROE case, we note that we should replace $\lambda(t)\rightarrow \lambda'(t)$, where $\lambda'(t) = \lambda(t)[\tm/(t+\tm)]^{2\alpha/3}$ in the time-integral in Eq.~\eqref{eq:ROE-sudakov}. Given the exponential decay of the Sudakov, in Eq.~\eqref{eq:rate-roe-limits}) one can simply substitute for $\omega_\text{\tiny BH}'(t) = \mu^2(t) \lambda(t)/2$ to obtain the desired effect. 

Similarly, a simple substitution of $\bar \omega_\text{c} \to \bar\omega_\text{c}'(t) = \mu^2(t) t/2$ can be done in Eq.~\eqref{eq:rate-oe-limits} for the OE case. The energy scale in the dilute regime is then  decaying slowly, $\bar\omega_c(t) \propto [\tm/(t+\tm)]^{2\alpha/3}t$, but this should be a small effect as long as $\tm$ is large compared to the initial mean-free-path $\lambda_0=\mu_0^2/\hat q_0$.  

For the IOE this modification results in a further logarithmic correction to $\hat q(t)$. The decaying screening mass will most significantly affect soft gluon emissions and we have estimated that the effective jet transport parameter receives a logarithmic correction term $\hat q_{\rm eff}(t) = \hat q(t) (1 + 0.646/\ln \frac{Q^2(t)}{\mu_\ast^2} -\frac{\alpha}{3}\ln\frac{\tm}{t+\tm}+ \ldots) $ followed by a revised minimal energy scale $\omega > 2\mu_\ast^4[\tm/(t+\tm)]^{4\alpha/3}\rme/\hat q_0(t)$, which is consistent with the decaying Bethe-Heitler energy, $\omega_\text{\tiny BH}'(t) \propto [\tm/(t+\tm)]^{4\alpha/3}$.
Note, however, that in some cases the medium may dilute rapidly enough to never enter the multiple-scattering regime. Apart from this particular case, for larger gluon energies we expect the corrections to be small, e.g. note that the $\omega \gg \omega_c(t)$ behavior in \eqref{eq:IOE-rate-hard} is insensitive to the screening mass. 

\section{Phenomenological applications}

\begin{figure*}[t!]
   \centering
   \includegraphics[width=0.9\textwidth]{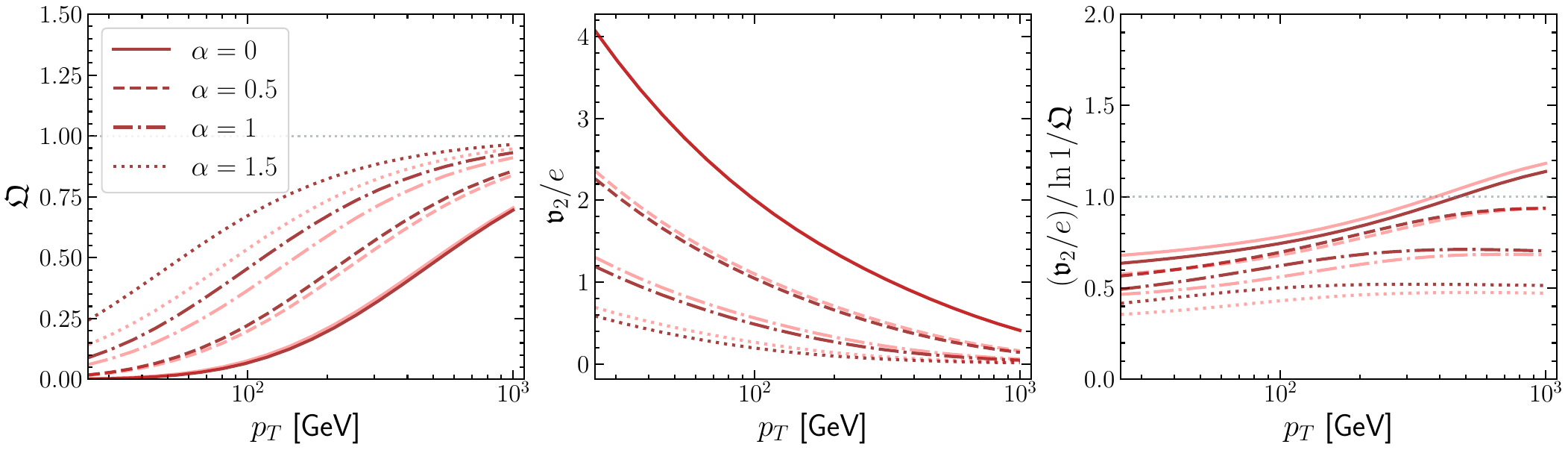}
   \caption{Enhancement of jet $v_2$ at fixed parton suppression factor between model \texttt{(ii)} and \texttt{(i)}. Curves from bottom to top correspond to different expansion with $0\leq \alpha \leq 1.5$. The rightmost figure reports the ratio of the respective values found in the two other columns.}
   \label{fig:fig3}
\end{figure*}

The medium-induced spectrum gives rise to energy loss which affects the overall suppression of high-$p_T$ particles or jets, quantified through the nuclear suppression factor $R_{AA}$, and to the azimuthal asymmetry of this suppression, quantified through the $v_2$ coefficient. 
To address the question that motivated our work, we propose an experimental observable that is sensitive to the medium's evolution during its earliest stages.

To simplify the analysis we consider fully coherent jets, i.e. jets whose opening angle is smaller than the critical resolution angle in the medium \cite{Casalderrey-Solana:2012evi}. In this case, the medium interactions can only resolve the total color charge of the jet and energy loss proceeds as off a single parton. This occurs for semi-central collisions at RHIC and LHC where the critical angle can be sizable \cite{Mehtar-Tani:2021fud,Mehtar-Tani:2024jtd}. To make this distinction clear in what follows, we introduce the single-parton quenching factor $\mathfrak{Q}(\pT)$ as a \textit{proxy} for the full jet suppression factor $R_{AA}(\pT)$. Similarly, the azimuthal asymmetry of the quenching for a single parton is denoted $\mathfrak{v}_2(\pT)$ as a \textit{proxy} for the full jet $v_2(\pT)$ measure. 

The quenching factor for a single parton due to radiative energy loss, which basically is the ratio of jet spectra in the presence and absence of medium effects, is well approximated by 
\begin{equation}
     \mathfrak{Q}(\pT) \simeq \exp\left[ -\int_0^L \rmd t \int_{\pT/n}^\infty \rmd \omega  \, \Gamma(\omega,t) \right] \,,
\end{equation}
where $n$ is the slope of the hard spectrum and $L$ is the medium length \cite{Baier:2001yt}. The azimuthal asymmetry of the quenching, for semi-central collisions, is readily found to be \cite{Mehtar-Tani:2024jtd}
\begin{equation}
    \frac{\mathfrak{v}_2 (\pT)}{e} \simeq - \frac12 \frac{\rmd \ln \mathfrak{Q}(\pT)}{\rmd \ln L } = \frac12 L \int_{p_T/n}^\infty \rmd \omega\, \Gamma(\omega,L) \,,
\end{equation}
where $e$ is the ellipticity of the nuclear overlap. While the quenching factor is sensitive to the accumulation of emissions along the entire in-medium path length $L$, the $\mathfrak{v}_2$ coefficient is directly sensitive to the rate at late times.

We will consider $L=5$ fm as a reference and we use the same parameters as in Sec.~\ref{sec:rates}, namely. For high-energy  collisions, where typically $n\gtrsim 5.5$, this immediately indicates that the Bethe-Heitler regime is not relevant for jet quenching at $\pT \gtrsim 10$ GeV. We plot a set of predictions for $\mathfrak{Q}(\pT)$ and $\mathfrak{v}_2(\pT)/e$ in Fig.~\ref{fig:fig3} for different medium expansion parameter $\alpha$. We have evaluated numerically the quenching factors $\mathfrak{Q}(\pT)$ and azimuthal coefficients $\mathfrak{v}_2/e$ in Fig.~\ref{fig:fig3} for the full, interpolated rate valid in all regimes for the same set of parameters as used above (the plotted observables are the nuclear suppression factor $\mathfrak{Q}$ (leftmost panel), the azimuthal coefficient $\mathfrak{v}/e$ (center panel) and their ratio (rightmost panel)). For different medium-evolution scenarios a weak $\pT$-dependence appears due to the evolving separation scale $\omega_c(t)$, and we observe that this can affect the relative weight of accumulated and late quenching as contained in $\ln 1/\mathfrak{Q}$ and $\mathfrak{v}_2/e$, respectively.

\begin{figure*}
    \centering
    \includegraphics[width=0.3\textwidth]{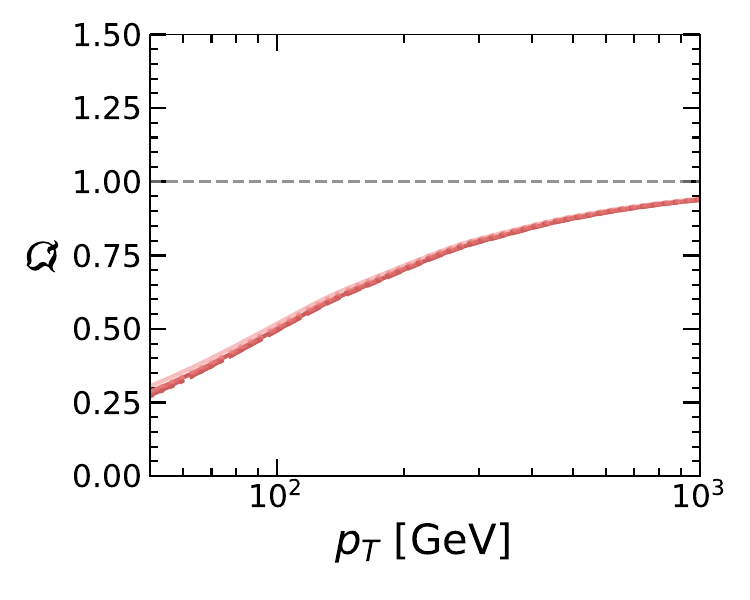}\includegraphics[width=0.3\textwidth]{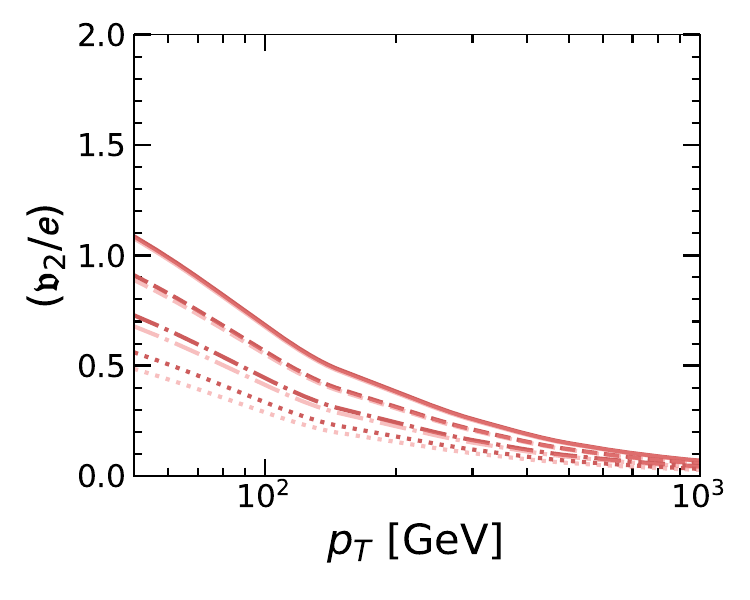}\includegraphics[width=0.3\textwidth]{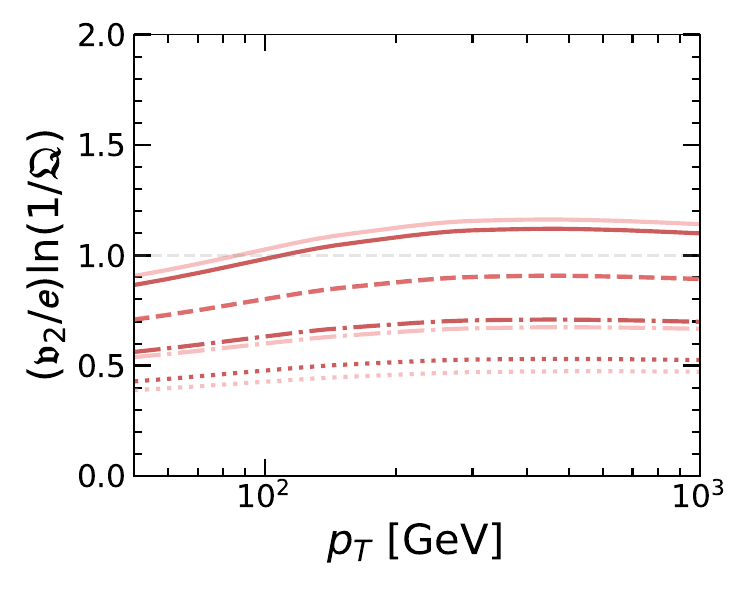}\\
    \includegraphics[width=0.3\textwidth]{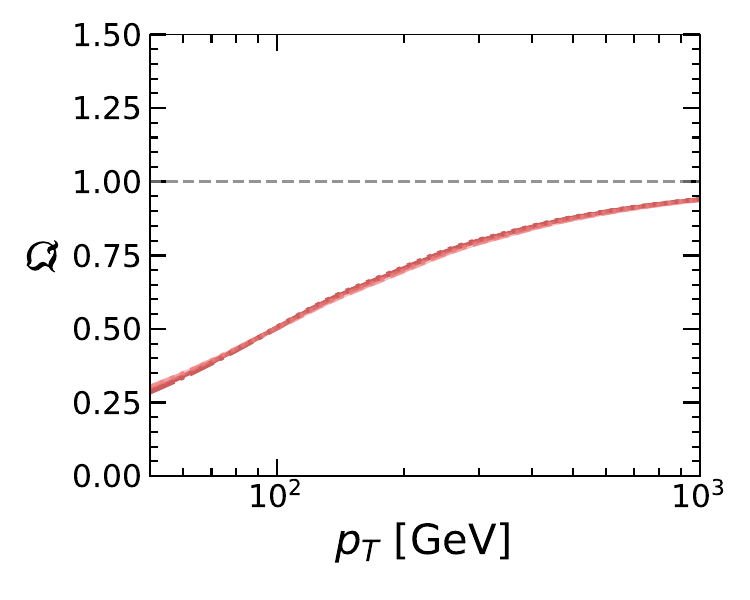}\includegraphics[width=0.3\textwidth]{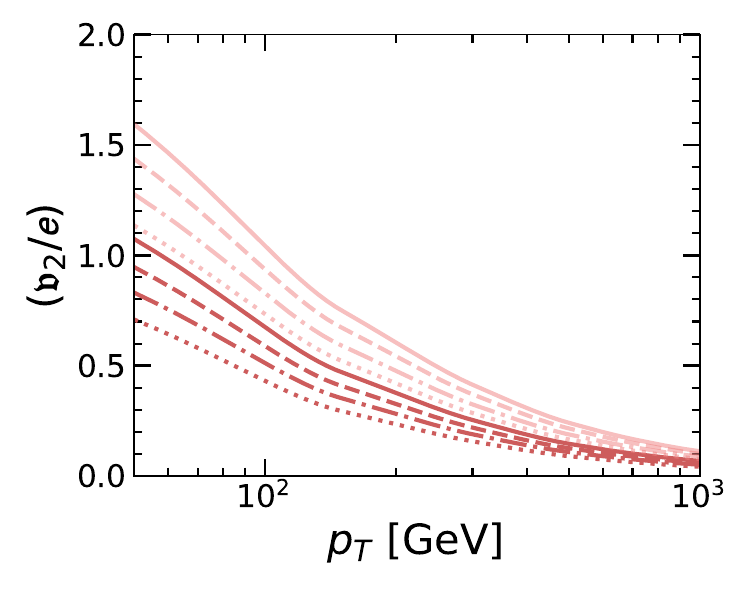}\includegraphics[width=0.3\textwidth]{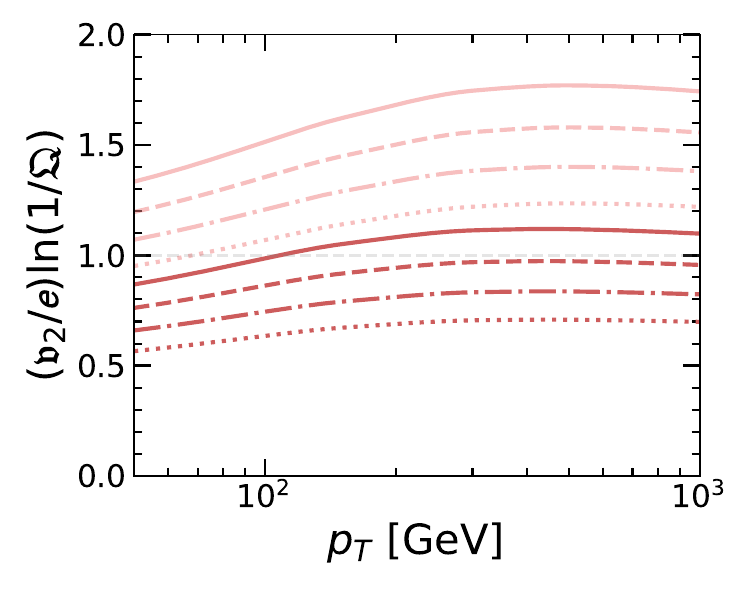}
     \caption{All reference scaling for $Q_{AA}$ at $\sim 0.5$ for $\pT =$ 100 GeV. Upper and lower panels for $\tm =1$ fm and $\tm=3$ fm, respectively.}.
     \label{fig:scaling}
\end{figure*}

In the following, we strive at establishing a relation between these two observables to clarify the role of the medium expansion scenarios $\texttt{(i)}$ versus $\texttt{(ii)}$.
Before solving the full numerical rates, we can gain analytical insights based on well founded approximations. Since emissions in the dilute phase or in the high-$\omega$ tail do not contribute significantly to quenching, i.e. $\mathfrak{Q}(\pT) \sim 1- \mathcal{O}(\alpha_s)$, we can neglect them altogether and focus emissions belonging to the multiple-scattering regime, see Eq.~\eqref{eq:rate-ioe-soft}, which lead to a sizable suppression factors $\mathfrak{Q}(\pT) \ll 1$ due to the high multiplicity of emissions \cite{Baier:2001yt}, see also \cite{Isaksen:2022pkj}. Considering the two models of expanding media defined in Eq.~\eqref{eq:qhat-models}, and using the simplified expressions for the soft-gluon rates, we readily obtain
\begin{align}
\label{eq:v2qaaratio}
    \frac{\mathfrak{v}^\text{\tiny soft}_2}{e} = \ln \frac{1}{\mathfrak{Q}^\text{\tiny soft}} \begin{cases} \frac{[x_m/(1+x_m)]^{\alpha/2}}{\Phi_1(x_m;\alpha/2)} & \text{for model \texttt{(i)}} \\ \frac{x_m^{\alpha/2}}{\Phi_2(x_m;\alpha/2)} & \text{for model \texttt{(ii)}} \end{cases} \,.
\end{align}
where $x_m \equiv \tm/L$ and we have neglected slow, logarithmic contributions and evaluate the ``dressed'' $\hat q$ at $L$. For scenario \texttt{(i)}, we get $\Phi_1(x;\alpha) = \frac{x^{\alpha}}{1-\alpha} \big((1+x)^{1-\alpha} - x^{1-\alpha} \big)$ while for scenario \texttt{(ii)} we arrive at $\Phi_2(x;\alpha) = \frac{x^{\alpha}}{1-\alpha} \big(1 - x^{1-\alpha} \big)$.

Remarkably, the enhancement factor on the right-hand side (within the curly bracket), obtained within these approximations, is $\pT$-independent. As a consequence, by fixing the quenching factor $\mathfrak{Q}(\pT)$, we deduce that the azimuthal asymmetry in model \texttt{(i)} will always be bigger than in model \texttt{(ii)}, i.e. $\mathfrak{v}_2^\texttt{(i)} < \mathfrak{v}_2^\texttt{(ii)}$, for any $0<x_m<1$ and for any $\alpha \geq 0$. In fact, the bigger the initial ``hydrodynamization'' time $\tm$, the bigger the difference.

These conclusions carry over to our final results that are obtained by using the full, numerical inelastic rates and presented in Fig.~\ref{fig:scaling}. We have considered models \texttt{(i)} and \texttt{(ii)} for different expansion parameters $0 \leq \alpha \leq 1.5$. We have also considered two representative values for $\tm$: a case of ``early" thermalization $\tm = 1$ fm and a case of ``late'' thermalization $\tm = 3$ fm. For a fair comparison and without loss of generality, we constrain the suppression factor at a single $\pT$ value $\mathfrak{Q}(\pT=100 \text{ GeV}) = 0.5$ in order to fix the jet quenching parameter at initial time, i.e. $\hat q_0(0)$. This reflects the typical suppression found in data, though specific experimental $R_{AA}(\pT)$ values for jets or hadrons at a given centrality could serve as an alternative baseline. The remaining parameters are kept to their default values, as above. The resulting values of $\hat q_0$ are summarized in Tab.~\ref{tab:qhat}.

Notably, the leftmost column of Fig.~\ref{fig:scaling} demonstrates that all medium scenarios predict nearly identical suppression factors, $\mathfrak{Q}(\pT)$, following a proper rescaling of the jet quenching parameter. However, the azimuthal asymmetry, quantified by $\mathfrak{v}_2(\pT)/e$, is more sensitive to the expansion parameter $\alpha$ (see Fig.~\ref{fig:scaling}, middle column). The resulting ratio $(\mathfrak{v}_2/e)/\ln(1/\mathfrak{Q})$ is nearly independent of $\pT$, consistent with the prediction in Eq.~\eqref{eq:v2qaaratio}.

For ``early'' thermalization ($\tm \leq 1$ fm), however, the two initial state models \texttt{(i)} and \texttt{(ii)} are indistinguishable. We observe a significantly larger $\mathfrak{v}_2$ coefficient for scenario \texttt{(ii)} only when the thermalization time is large relative to the medium length ($\tm = 3$ fm), as shown in the lower right panel of Fig.~\ref{fig:scaling}. This demonstrates that a combined analysis of overall jet suppression and azimuthal anisotropy offers sensitivity to the details of early jet quenching.

\begin{table}[]
    \begin{tabular}{|cccccc|}
    \hline
    \multicolumn{1}{|c|}{$ \tm $} & \multicolumn{1}{c|}{$\hat{q}_0(0)$ [GeV$^3$]} & \multicolumn{1}{c|}{$\alpha$ = 0} & \multicolumn{1}{c|}{$\alpha$ = 0.5} & \multicolumn{1}{c|}{$\alpha$ = 1.0} & $\alpha$ = 1.5 \\ \hline
    \multicolumn{1}{|c|}{\multirow{2}{*}{ 1 fm }} & \multicolumn{1}{c|}{model \texttt{(i)}} & \multicolumn{1}{c|}{0.068} & \multicolumn{1}{c|}{0.14} & \multicolumn{1}{c|}{0.27} & 0.5 \\ \cline{2-6} 
    \multicolumn{1}{|c|}{} & \multicolumn{1}{c|}{model \texttt{(ii)}} & \multicolumn{1}{c|}{0.069} & \multicolumn{1}{c|}{0.125} & \multicolumn{1}{c|}{0.21} & 0.33 \\ \hline
    \multicolumn{1}{|c|}{\multirow{2}{*}{ 3 fm }} & \multicolumn{1}{c|}{model \texttt{(i)}} & \multicolumn{1}{c|}{0.068} & \multicolumn{1}{c|}{0.097} & \multicolumn{1}{c|}{0.138} & 0.19 \\ \cline{2-6} 
    \multicolumn{1}{|c|}{} & \multicolumn{1}{c|}{model \texttt{(ii)}} & \multicolumn{1}{c|}{0.106} & \multicolumn{1}{c|}{0.123} & \multicolumn{1}{c|}{0.14} & 0.16 \\ \hline
    \end{tabular}
\caption{The values of the initial $\hat{q}_0$ for models \texttt{(i)} and \texttt{(ii)}, and for $0 \leq \alpha \leq 1.5$, found from fixing $\mathfrak{Q}(\pT= 100 \text{ GeV}) = 0.5$. We have also considered two ``hydronamization'' times: ``early" $\tm = 1$ fm and ``late'' $\tm = 3$ fm.}
\label{tab:qhat}
\end{table}

\section{Conclusions}
\label{sec:conclusions}

Jet suppression results from an accumulation of processes, spanning from the initial creation of a hard parton in the earliest stages of a hard collision to the final hadronization and particle detection. The evolution of the jet is intimately tied to that of the medium; however, accessing the details of the earliest---and arguably most interesting---stages requires controlling for many confounding factors and a rigorous phenomenological analysis of existing observables.

In this work, we have provided a set of novel rates applicable to any medium expansion scenario. These rates are (semi-)analytical, valid across the full phase space of gluon energies $\omega$ at any given time $t$ during passage through the medium, and carefully account for the regime of multiple scattering in expanding media. They can thus serve as important inputs for future phenomenological analyses.

Using these rates within a simplified medium setup, we present predictions for the suppression factor of coherent jets (in the single-parton approximation) and their azimuthal asymmetry, quantified by $\mathfrak{Q}(\pT)$ and $\mathfrak{v}_2(\pT)/e$, respectively. While the overall suppression is sensitive only to the total magnitude of quenching---which can be reabsorbed into a redefinition of the initial medium parameters---the combination of these two observables provides insight into the details of the earliest stages. The distinction between these observables, arising from geometry and the resulting evolving momentum scales, could become an important tool to pin down the early evolution of $\hat q$ in heavy-ion collisions.

We expect further progress to be made within a more sophisticated approach that accounts for medium geometry more realistically. In such a framework, the expansion histories of the medium along jet trajectories could be sampled using viscous hydrodynamics models matched onto ``attractor'' solutions for the early stages of heavy-ion collisions, see e.g.~\cite{Pablos:2025cli}.
\newline
\paragraph*{Acknowledgements:} SPA acknowledges funding from grant agreement PAN.BFD.S.BDN.612.022.2021-PASIFIC 1, QGPAnatomy. This work received funding from the European Union’s Horizon 2020 research and innovation program under the Maria Sklodowska-Curie grant agreement No.\ 847639, the Polish Ministry of Education, the Horizon Europe Programme during the initial stages of work. SPA is supported
by the Marie Sklodowska-Curie Actions – COFUND project, which is co-funded by the European
Union (Physics for Future – Grant Agreement No. 101081515. SPA also acknowledges Department of Theoretical Physics, CERN (CERN-TH visitor program) where this work was initiated. 
\appendix
\section{Solutions of harmonic equation}
\label{sec:IOE-solutions}

The expansion of the full 3-point function in transverse-coordinate space $\Kc(\x,\y)$ can be cast as an implicit equation,
\begin{equation}
    \begin{split}
    \label{eq:cK_ful_IOE_app}
        &\cK(\x,t_2;\y,t_1) = \cK_\HO(\x,t_2;\y,t_1) \\
        &-\int_{t_1}^{t_2}  \rmd s \int_\z \, \cK_\HO(\x,t_2;\z,s) \delta v(\z,s)\cK(\z,s;\y,t_1) \,,  
    \end{split}
\end{equation}
where $\Kc_\HO(\x,\y)$ is the full correlator with the harmonic potential $v_\HO(\x,t)$. The exact solution then reads \cite{Kleinert:2004ev,Arnold:2008iy},
\begin{align}
\label{eq:cK_BDMPS_app}
    &\cK_ {\rm HO}(\x,t_2;\y,t_1)= \frac{\omega}{2\pi i\, S(t_2,t_1)} \exp\Bigg\{ \frac{i\omega}{2S(t_2,t_1)} \nn
    &\times\bigg[ C(t_1,t_2)\,\x^2+C(t_2,t_1)\,\y^2-2 \x\cdot\y\bigg] \Bigg\} \,.
\end{align}
The functions $C(t,t_0)$ and $S(t,t_0)$ both satisfy the harmonic equation $\big[\partial_t^2 + \Omega^2(t) \big] f(t,t_0) =0$, where $\Omega^2(t) = -i \hat q(t)/(2\omega)$. The two sets of different boundary conditions are $S(t_0,t_0)=0$ and $\partial_t  S(t,t_0)|_{t=t_0} = 1$ and $C(t_0,t_0)=1$ and $\partial_t C(t,t_0)|_{t=t_0}=0$, respectively. These solutions are related by the associated Wronskian, which reads $ W = C(t,t_0) \partial_{t}S(t,t_0) -S(t,t_0)\partial_{t} C(t,t_0) = 1$ for the above initial conditions. This implies that $S(t,t_0) = -S(t_0,t)$. 

\paragraph{Vacuum solutions:} The vacuum solutions for the two functions are enforced by the boundary conditions and are simply $S(t,t_0) = t-t_0$ and $C(t,t_0)=1$.

\paragraph{Medium with constant density:} With a constant density, $\hat q(t) = \hat q$, the solutions are well known $S(t,t_0) = \sin\big[\Omega(t-t_0) \big]/\Omega$ and $C(t,t_0) = \cos \big[\Omega(t-t_0)\big]$, where $\Omega = \sqrt{-i\hat q/(2\omega)}$.

\paragraph{Expanding medium (scenario 1):} The first scenario suggested in the paper, proposes a medium that continuously is decaying from a constant value, namely $\hat q(t) = \hat q\, [\tm/(t+\tm)]^\alpha$. Defining $\nu=1/(2-\alpha)$ and introducing the combined variable $z_t = 2 i \nu \Omega \, \tm[(t+\tm)/\tm]^{\frac{1}{2\nu}}$, we can write them as
\begin{align}
\label{eq:S-expanding}
    S(t,t_0) &= 2\nu\,\sqrt{(t+t_m) (t_0+t_m)} \nonumber\\ &\times \Big[I_\nu(z_t) K_\nu(z_{t_0}) - I_\nu(z_{t_0}) K_\nu(z_t) \Big] \,,\\
\label{eq:C-expanding}
    C(t,t_0) &= 2 i \nu \Omega \, \sqrt{t+\tm}(t_0+\tm)^\frac{1-\nu}{2\nu} \nonumber\\ &\times \Big[I_\nu(z_t) K_{\nu-1}(z_{t_0}) + I_{\nu-1}(z_{t_0}) K_\nu(z_t) \Big] \,.
\end{align}

\paragraph{Expanding medium (scenario 2):} The second scenario suggested in the paper, consists of a vacuum-like phase up to a medium thermalization time $\tm$, i.e. $\hat q(t) = 0$ for $t<\tm$, and a decaying medium afterwards, i.e. $\hat q(t) = \hat q\, (\tm/t)^\alpha$ for $t\geq \tm$. Since for a time ordered $t > t_1>t_0$, any solution in $(t_2,t_1)$ can be written as~\cite{Arnold:2008iy}, $S(t,t_0) =C(t_0,t_1)S(t,t_1)-S(t_0,t_1)C(t,t_1)$ and $C(t,t_0) =-\partial_{t_0}C(t_0,t_1)S(t,t_1)+\partial_{t_0}S(t_0,t_1)C(t,t_1) $. By assigning the intermediate time $t_1$ to $\tm$, and utilizing the vacuum solutions for the $(0,\tm)$ interval, we find
\begin{align}
S(t,0) &= \tilde S(t,\tm) + \tm \tilde C(t,\tm) \,,\\
C(0,t) &= \tilde C(\tm,t) + \tm \partial_t \tilde C(t,\tm) \,,
\end{align}
where $\tilde S(t,t_0)$ and $\tilde C(t,t_0)$ can be read off Eqs.~\eqref{eq:S-expanding} and \eqref{eq:C-expanding} by substituting $t+\tm\to t$ and $t_0 +\tm \to t_0$. Hence, the corrections are of order $\tm$ and does not affect the soft behavior of the functions.

\paragraph{Generic solutions:} Assuming that the solutions to $S(t,t_0)$ and $C(t,t_0)$ are the linear combinations of two independent solutions to the second-order differential equations, i.e. $c_1 f_1(t) + c_2 f_2(t)$, we find
\begin{align}
    S(t,t_0) &= \frac1{\Phi(t_0)} \left(\frac{f_1(t)}{f_1(t_0)} - \frac{f_2(t)}{f_2(t_0)} \right) \,,\\
    C(t,t_0) &= \frac1{\Phi(t_0)} \Bigg(- [\partial_t\ln f_2(t)]_{t=t_0}\frac{f_1(t)}{f_1(t_0)} \nonumber\\ &\qquad +[\partial_t\ln f_1(t)]_{t=t_0} \frac{f_2(t)}{f_2(t_0)} \Bigg) \,,
\end{align}
where we defined $\Phi(t_0) =\left[\partial_t \ln f_1(t) - \partial_t \ln f_2(t) \right]_{t=t_0}$.
\bibliographystyle{apsrev4-1}
\bibliography{main}

\end{document}